\documentclass[a4paper,11pt]{article}
\usepackage{pos}
\usepackage{color}
\usepackage[numbers]{natbib}

\newcommand{\beq}{\begin{equation}}
\newcommand{\eeq}{\end{equation}}
\newcommand{\bea}{\begin{eqnarray}}
\newcommand{\eea}{\end{eqnarray}}

\title{Particle acceleration at the termination shock of stellar clusters' wind}
 \ShortTitle{Particle acceleration at wind of stellar clusters}

\author*[a]{Giovanni Morlino}

\affiliation[a]{INAF/Osservatorio Astrofisico di Arcetri,\\
  L.go E. Fermi 5, 50125, Firenze, Italy}

\emailAdd{giovanni.morlino@inaf.it}

\abstract{
We investigate the process of particle acceleration at the termination shock that develops in the bubble excavated by star clusters’ winds in the interstellar medium. We develop a theory of diffusive shock acceleration at such shock and we find that the maximum energy may reach the PeV region for very powerful clusters.
We show how the maximum energy is limited by two different processes: the particle escape from the bubble boundary and the drop of energy gain for particles able to diffuse up to the center of the cluster.
A crucial role in this problem is played by the dissipation of kinetic energy of the wind to magnetic perturbations which determines the diffusion regime of particles: in case the diffusion is close to Bohm than PeV energies can be reached. }

\FullConference{37$^{\rm{th}}$ International Cosmic Ray Conference (ICRC 2021)\\
		July 12th -- 23rd, 2021\\
		Online -- Berlin, Germany}

\begin{document}
\maketitle

\section{Introduction}
In recent years, massive stellar clusters (MSC) have received renewed attention as possible contributors to the Galactic component of cosmic rays (CR). This possibility was already pointed out in the '80s by Cesarsky \& Montmerle \cite{Cesarsky-Montmerle:1983}, where they show that the power released by stellar winds in the Galaxy is of the same order of the one released by SN explosions, whose remnants are considered as the prime candidates for the acceleration of Galactic CRs. However, contrary to supernova remnants (SNR) which are well known non-thermal emitters, a clear association between MSCs and $\gamma$-ray emission has been reported only recently \citep{Abramowski_Wd1:2012, Yang+2018, Ackermann:2011p3159,Aharonian+2019NatAs, Saha_NGC3603:2020}. Moreover, in at least one case, the Cygnus OB2 association, LHAASO detected photons up to 100 TeV \citep{LAHHSO2021Nature594}, suggesting the presence of PeV protons if the emission were of hadronic origin. Such a finding has triggered several speculations about particles acceleration in such sources.
The majority of proposed models assumes that particles are accelerated in the innermost region of the clusters, either by colliding stellar winds, or by the turbulence resulting from such collisions \cite[see][for a review]{Bykov+2020}.
A notable alternative to those models is the possibility to accelerate particles at the termination shock (TS) formed by the large scale wind collectively formed by the winds of many stars in the cluster \citep{Morlino+2021}. Such a large scale wind can be generated if the cluster is compact enough that the average distance between stars is smaller than the termination shock of a single wind \citep{Gupta+2018}. 

For many years the acceleration at the stellar wind termination shock has been disregarded for tho main reasons: on one hand the wind is never faster than $\sim 3000$\,km/s, hence quite smaller if compared to shocks in young SNRs. On the other hand, the magnetic field just upstream of the termination shock (hence in the free expanding region) should be quite small if resulting only from the stellar magnetic field adiabatically expanded. Those two conditions would imply a quite small maximum energy for accelerated particles. In addition, if the  magnetic field structure is like a Parker spiral, as expected for a wind from a single star, the injection of particle in the shock acceleration mechanism remains suppressed.

However, in a stellar cluster a strong magnetic turbulent may develop as a consequence of both wind-wind collisions and intermittency of the wind itself, allowing to reach a strength of several $\mu$G at the location of the TS. In addition, those systems have a long lifetime, of the order of several Myr, hence the maximum energy is presumably limited by their size, rather than by their age as happen for SNRs. Considering that the TS can be located at tens of pc from the cluster while the size of the hot bubble produced by the shock wind can reach hundreds of pc, one can speculate that very high energy may indeed be reached.

Another interesting consequence of particle acceleration at wind TS is the fact that the chemical composition of accelerated particles will reflect the wind composition which is different form the average ISM composition. This peculiarity may help solving some anomalies observed in the CR spectrum like the $^{22}$Ne excess \cite{Gupta+2020}.

In this work we present the first model for the particle acceleration at the wind termination shock in a spherical geometry. This model has been fully developed in \cite{Morlino+2021}, hence the reader may refer to such publication for further details.

\section{Bubble dynamics}
\label{sec:bubble}
\begin{figure}
\centering
\includegraphics[width=.35\textwidth]{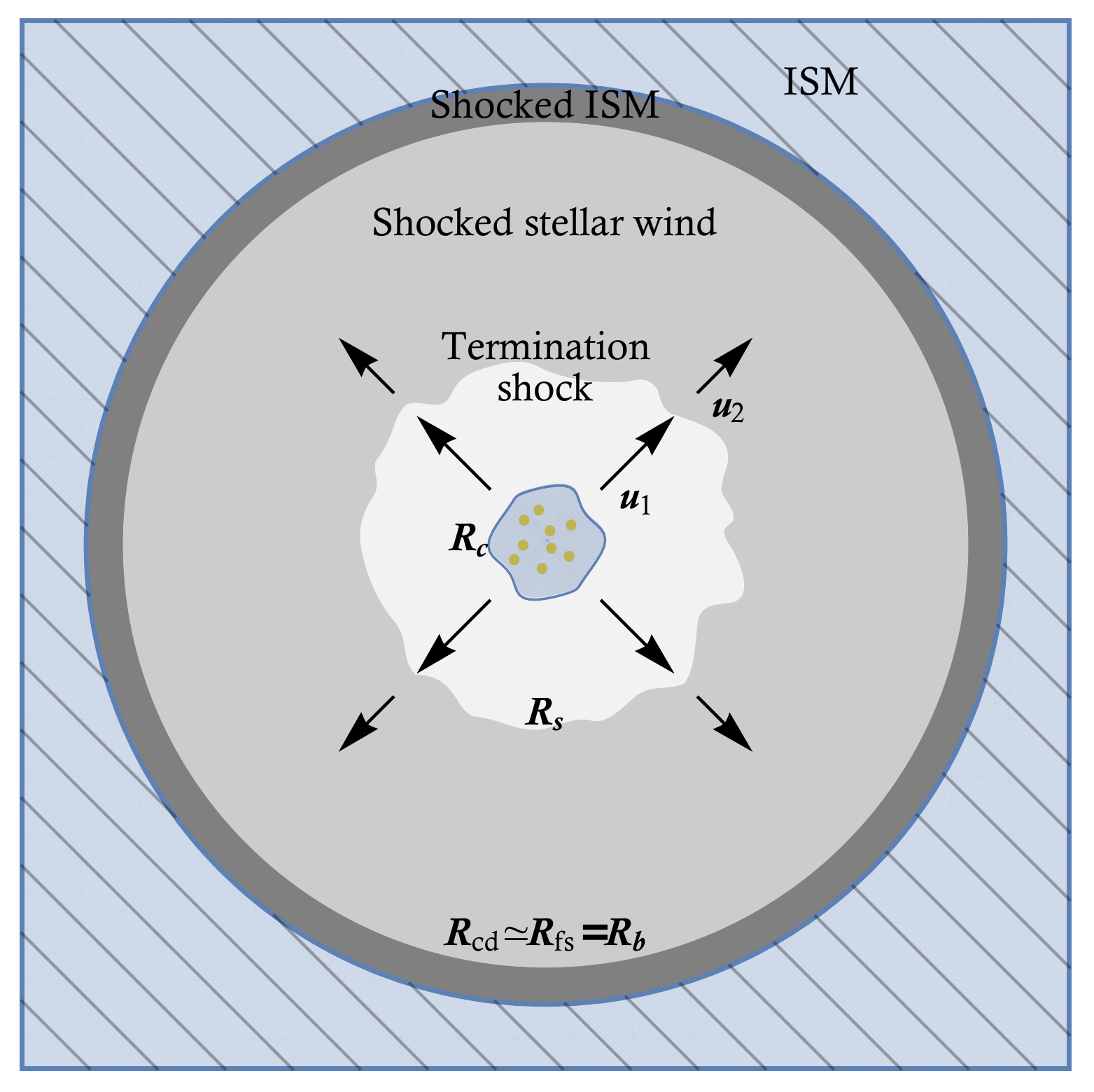}
\caption{Schematic structure of a wind bubble excavated by a star cluster into the ISM: $R_{s}$ marks the position of the termination shock, $R_{\rm cd}$ the contact discontinuity, and $R_{\rm fs}$ the forward shock.}
\label{fig:geometry}
\end{figure}
The bubble excavated by the collective stellar wind launched by the star cluster is schematically illustrated in Figure~\ref{fig:geometry}: the central part is filled with the wind itself, expanding with a velocity $v_{w}$ and density $
\rho(r) = \frac{\dot M}{4\pi r^{2} v_{w}}$ for $r > R_{c}$,
where $R_{c}$ is the radius of the core where the stars are concentrated, and $\dot M$ is the rate of mass loss due to the collective wind. The impact of the supersonic wind with the ISM, assumed here to have a constant density $\rho_{0}$, produces a forward shock at position $R_{\rm fs}$, while the shocked wind is bound by a termination shock, at a location $R_{s}$. The bubble spend the majority of its lifetime in a phase where the expansion is adiabatic, while the shocked ISM is compressed into a thin shell due to the cooling time much shorter than the system typical age, hence we can assume that the contact discontinuity is located at $R_{\rm cd}\approx R_{\rm fs} \equiv R_{b}$. In those conditions the bubble structure is determined only by the wind luminosity, $L_w= \frac{1}{2} \dot{M} v_w^2$, and the density of the external medium \citep{Weaver+1977}: the location of the TS and bubble boundary are
\begin{eqnarray}
    R_{s} = 48.6 ~ \dot{M}_{-4}^{3/10} v_{8}^{1/10} \rho_{1}^{-3/10} t_{10}^{2/5} ~\rm pc \, \\
    R_{b} = 220  ~ \dot{M}_{-4}^{1/5}  v_{8}^{2/5}  \rho_{1}^{-1/5}  t_{10}^{3/5} ~\rm pc \,,
\end{eqnarray}
where $v_{8}=v_{w}/(1000~\rm km\, s^{-1})$ and $\dot M_{-4}=\dot M/(10^{-4}\rm M_{\odot}\,yr^{-1})$, $\rho_{1}$ is the ISM density in the region around the star cluster in units of 1 proton per cm$^{3}$ and $t_{10}$ is the dynamical time in units of 10 million years. It is worth noticing that the shell moves outwards with a velocity $\dot R_b$ that is only a few tens of km/s, thereby being at most transonic. For this reason the forward shock should be unable to accelerate particles and will be neglected here.
The wind velocity in the shocked region can be derived assuming constant pressure in the hot bubble which implies $u(r>R_s)= v_w/\sigma (r/R_s)^{-2}$ where $\sigma$ is the compression ratio at the TS (equal to 4 for strong shocks).

The particle diffusion is determined by the structure of magnetic turbulence.
We assume here that a fraction $\eta_B$ of the wind kinetic energy is converted to magnetic turbulence at any location, in such a way that the strength of the turbulent magnetic field can be written as
$B(r) \approx \frac{1}{r}\left( \frac{1}{2}\eta_{B} \dot M v_{w} \right)^{1/2}.$
At the location of the termination shock, the strength of the magnetic field reads
\beq \label{eq:B_TS}
B(R_{s})=3.7\times 10^{-6} ~\eta_{B}^{1/2} \dot M_{-4}^{1/5}v_{8}^{2/5} \rho_{1}^{3/10} t_{10}^{-2/5}~ \mu \rm G.
\eeq
This dissipation of kinetic energy into magnetic energy likely results in turbulence with a typical scale $L_c$ that is expected to be of order the size of the star cluster, $L_c\sim R_c\sim 1\div 2$ pc. If the turbulence evolves following a Kolmogorov cascade, the diffusion coefficient immediately upstream of the termination shock can be estimated as
\begin{flalign}  \label{eq:DKol}
 D(E)\approx\frac{1}{3} r_{L}(p) v \left(\frac{r_{L}(p)}{L_{c}}\right)^{-2/3} = 
 2\times 10^{26} 
 \left( \frac{L_c}{1\rm pc}\right)^{2/3} 
  \eta_{B}^{-1/6} \dot M_{-4}^{-1/15} v_{8}^{-2/15} \rho_{1}^{-1/10} t_{10}^{2/15} E_{\rm GeV}^{1/3}~\rm cm^2~s^{-1},
\end{flalign}
where $r_{L}(p)=pc/e B(r)$ is the Larmor radius of particles of momentum $p$ in the magnetic field $B(r)$. The diffusion coefficient decreases inward as $(r/R_{s})^{1/3}$. One can see that the dependence of the diffusion coefficient upon the efficiency of conversion of kinetic energy to magnetic energy, $\eta_{B}$, is very weak. Downstream of the termination shock, it is assumed that the magnetic field is only compressed by the standard factor $\sqrt{11}$, typical of a strong shock, so that $D_{2}\approx 0.67 D_{1}$. Clearly the downstream diffusion coefficient can be smaller than this estimate suggests, if other processes (such as the Richtmyer-Meshkov instability \cite[]{Giacalone:2007p962}) lead to enhanced turbulence behind the shock.

\section{Model for particle acceleration}
\label{sec:acceleration}
With respect to SNRs, the geometry of wind-blown bubble is reversed, in the sense that the upstream is located in the inner part of the system (unshocked wind), hence the escape of particles can only occurs from the bubble boundary, downstream of the shock. Such a peculiar geometry requires a detailed calculation of the spectrum and spatial distribution of the particles accelerated at the termination shock. Given the quasi-stationary evolution of the wind region, the CR transport is modelled using the standard time-independent transport equation in spherical symmetry: 
\beq
  \frac{\partial}{\partial r} \left[ r^2 D(r,p) \frac{\partial f}{\partial r} \right]
  - r^2 u(r) \frac{\partial f}{\partial r} 
  + \frac{d\left[ r^2 u(r) \right]}{d r} \frac{p}{3} \frac{\partial f}{\partial p} 
  + \, r^2 Q(z,p) = 0  \,,
  \label{eq:transport}
\eeq
where $u(r)$ is the plasma speed and $D(r,p)$ is the diffusion coefficient. 
Particle acceleration takes place only at the termination shock, located at $r=R_s$, where particles are injected according to:
\beq
  Q(r,p) = Q_0(p) \delta(r-R_{s}) 
            = \frac{\eta_{\rm inj }n_{1} u_1}{4 \pi p_{\rm inj}^2} \delta(p-p_{\rm inj}) \delta(r-R_{s}),
  \label{eq:Qinj}
\eeq
where $n_{1}$ is the density of the cold wind immediately upstream of the termination shock and $\eta_{\rm inj}$ is the fraction of particle flux that takes part in the acceleration process. 

The solution of the transport equation is found in three steps: first solving the equation upstream ($f_1$) and downstream ($f_2$) and then in the region outside of the bubble ($f_3$) where $u=0$ and the diffusion reduces to the average Galactic one, i.e. $D=D_{\rm gal}$. Finally, using the flux conservation, the three solutions are joined together at the shock surface, where the solution is $f_s(p)\equiv f(r=R_s,p)$, and at the bubble boundary, where $f_b(p)\equiv f(r=R_b,p)$.
We also need to specify two boundary conditions: one is fixed assuming no net flux at $r=0$, namely $\left[r^2 D\partial_r f - r^2 u f \right]_{r=0}=0$, while the second one requires than the solution match the Galactic CR distribution at infinity, i.e. $f(r=\infty,p) = f_{\rm gal}(p)$. Notice that this later boundary condition is different from what we have done in \citep{Morlino+2021}, where we assumed $f_{\rm gal}= 0$ for simplicity.
Because of the spatial dependence of the plasma velocity and the spherical symmetry, the solution is found adopting an iterative procedure, similar to that introduced to treat non-linear DSA \cite[see, e.g.][]{Amato-Blasi2006}.

The spatial profile of the solution in the three different regions reads as follow:
\bea
  f_1(r,p) &=& f_s(p) \, \exp { \left\{-\int_r^{R_s} \alpha_1 \left[ 1+ \frac{R_s^2 \Lambda_1(r',p)}{r'^2 f_1(r',p)} \right] \frac{dr'}{R_s} \right\} } \,.  \label{eq:f1}\\
  f_2(r,p) &=& f_{s}(p) \, e^{\alpha_2(r)} \frac{1+\beta[e^{\alpha_2(R_b)}e^{-\alpha_2(r)}-1]}{1+\beta[e^{\alpha_2(R_b)}-1]} 
    + f_{\rm gal}(p) \, \frac{\beta[e^{\alpha_2(r)}-1]}{1+\beta[e^{\alpha_2(R_b)}-1]}  \label{eq:f2} \\
  f_3(r,p) &=& f_b(p) \frac{R_b}{r} + f_{\rm gal}(p) \left( 1 - \frac{R_b}{r} \right)  \label{eq:f3}
\eea
where
\begin{equation*}
\alpha_1(r,p)= \frac{u_1 R_s}{D_1(r,p)} \;,\; 
\alpha_2(r,p)= \frac{u_2}{R_{s} D_2(p)} \left(1 - \frac{R_{s}}{r} \right) \; , \;  
\beta(p)= \frac{D_{\rm gal}(p) R_{s}}{u_2 R_b^2} \,
\end{equation*}
\beq  \label{eq:G1_simply}
  \Lambda_1(\xi,p) =  -\frac{2}{3} \int_{0}^{\xi} f_1(\xi,p) \frac{d \ln(p^3 f_1)}{d \ln p}  \xi' d\xi' ,
\eeq
with $\xi \equiv r/R_s$. Notice that Eq.~\eqref{eq:f1} is non linear because it depends of $f_1$ through $\Lambda_1$. To find the full solution we have to explicit also the distribution function at the termination shock which reads
\beq	\label{eq:fs}
  f_s(p) = s k \left( \frac{p}{p_{\rm inj}} \right)^{-s}  e^{ -\Gamma_1(p)} e^{ -\Gamma_2(p)} \,, 		
\eeq
where $k=\eta_{\rm inj} n_1/(4\pi p_{\rm inj}^3)$ and $s= 3 u_1/(u_1-u_2)$.
Notice that $\eta_{\rm inj}$ snd $p_{\rm inj}$ are connected to the CR acceleration efficiency, $\epsilon_{\rm CR}$, defined as the fraction of wind kinetic luminosity converted into CRs, namely $\epsilon_{\rm CR} L_w = L_{\rm CR} \equiv 4\pi R_{s}^2 u_2 \int f_{s}(p) E(p) \, d^3p$. 

Eq.~\eqref{eq:fs} contains three terms. The first one is the usual power law $\propto p^{-s}$ that one finds in the plane parallel shock case while the two exponential terms are 
\bea
  \Gamma_1(p) = s \int_{p_{\rm inj},}^{p} \frac{\Lambda_1(R_s,p')}{f_s(p')} \, \frac{dp'}{p'}  \,, \label{eq:Gamma1} \\
  \Gamma_2(p) = \frac{s}{\sigma} \int_{p_{\rm inj},}^{p} \frac{1}{e^{\alpha_2(p',R_b)} - 1} \, \frac{dp'}{p'}  \,. \label{eq:Gamma2}
\eea
The former, $\Gamma_1$, depends on both $f$ and $f_s$. Hence, the solution can be obtained using an iterative methods that solves simultaneously Eqs.~\eqref{eq:f1} and \eqref{eq:fs}. Once the convergence is reached, Eqs.~\eqref{eq:f2} and \eqref{eq:f3} are also uniquely determined.

\begin{figure}
\centering
\includegraphics[width=.49\textwidth]{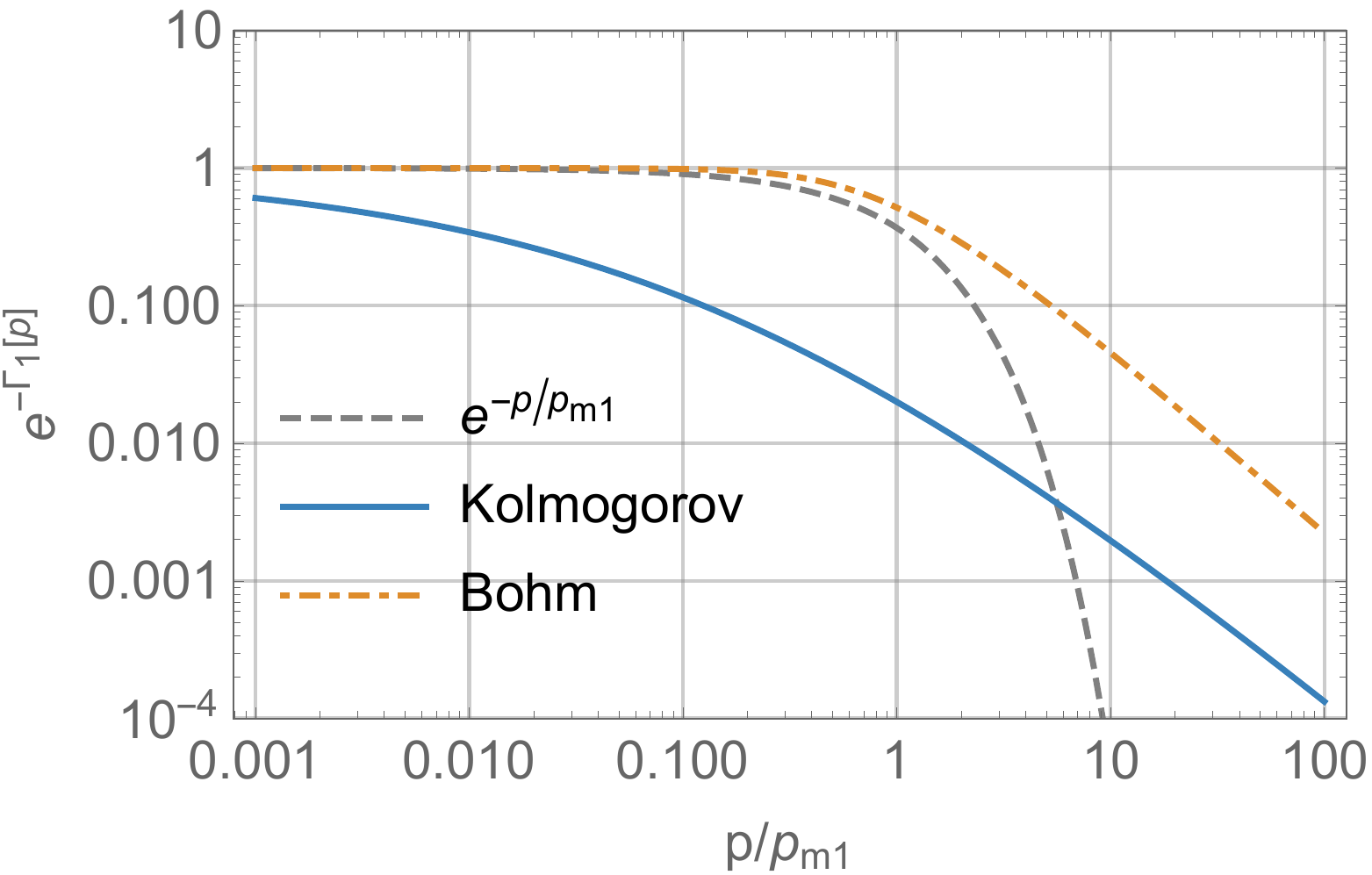}
\includegraphics[width=.49\textwidth]{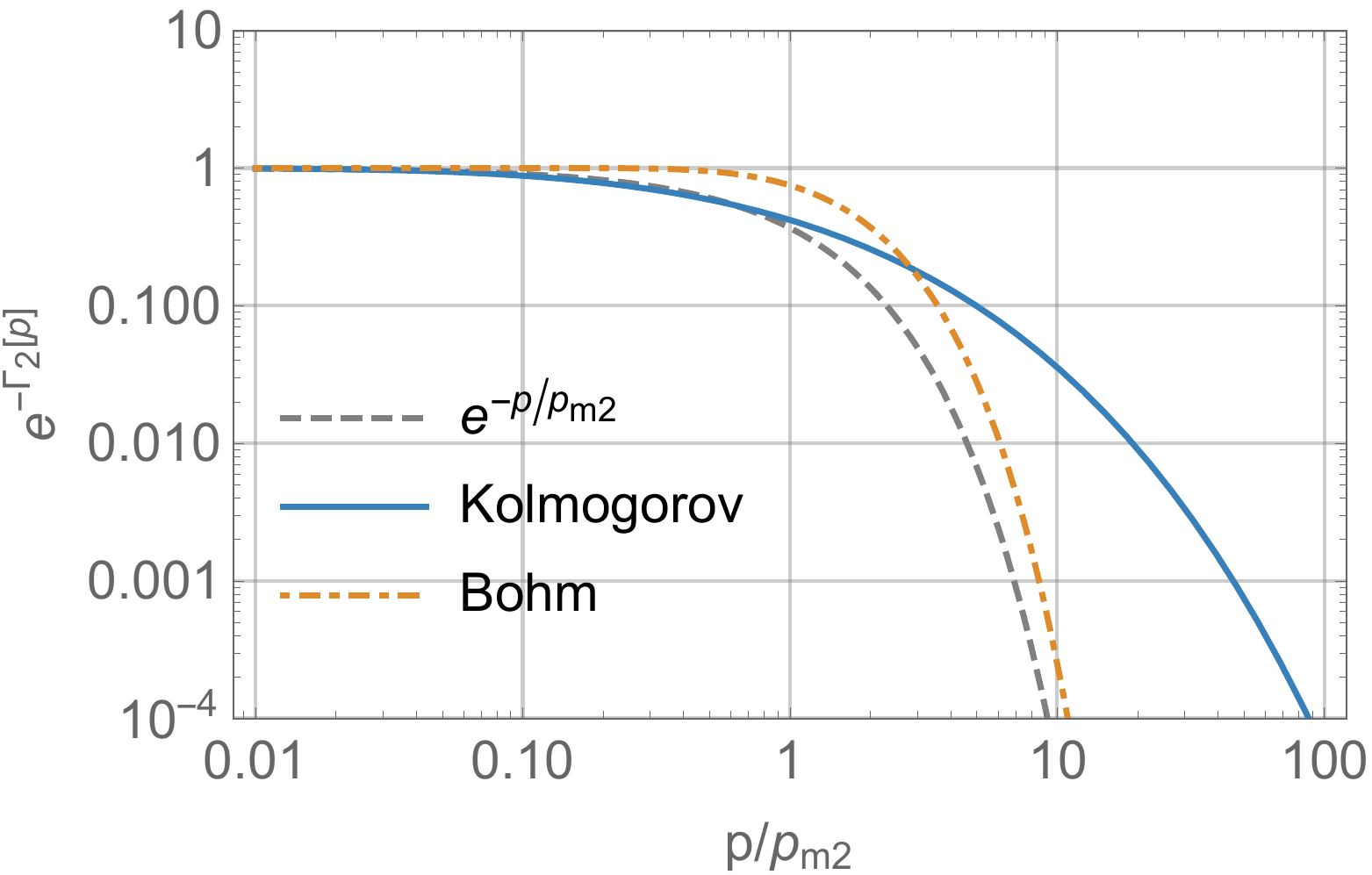}
\caption{Exponential functions $e^{-\Gamma_1(p)}$ (upper panel) and $e^{-\Gamma_2(p)}$ (lower panel) for the case of Kolmogorov and Bohm diffusion. For comparison the simple exponential function is also shown (gray dashed line).}
\label{fig:cutoff}
\end{figure}

The functions $\Gamma_1$ and $\Gamma_2$ are both positively defined and contain information about the geometry of the system, therefore determining the maximum energy of accelerated particles as we are going to discuss.
The physical meaning of the suppression due to $\Gamma_1$ can be understood in terms of particle energy gain. For plane parallel shocks, in the test particle limit, the energy gain per cycle is given by $\Delta E/E =4 (u_1-u_2)/(3c)$. In a more general approach $u_1$ and $u_2$ should be replaced by the effective velocities felt by particles upstream, $u_{p1}$, and downstream, $u_{p2}$. In a spherical geometry the effective velocity in the upstream can be written as \citep{Berezhko-Voelk:1997}:
\beq
 u_{p1} = u_1 - \int_{0}^{R_s} dr \frac{\partial(r^2 u)}{\partial r} \frac{f(r,p)}{f_s(p) R_s^2} \,.
 \label{eq:u_p}
\eeq
while $u_{p2} = u_2$ because in the downstream $\partial_r (r^2 u) =0$. 
We can now define the characteristic momentum $p_{\rm m1}$ as the one where particles can diffuse effectively up to the cluster centre, namely when $D_1/u_1 = R_s$ (or $\alpha_1=1$).
Using the approximate expression for the distribution function upstream at the first order, namely $f_1(\xi,p) = f_s(p) \exp\left[-(1-\xi) \alpha_1 \right]$, it is easy to see that for $p\gg p_{\rm m1} \Rightarrow u_{p1} \simeq u_1 \alpha_1/3$, while for for $p\ll p_{\rm m1} \Rightarrow u_{p1} \simeq u_1(1-2/\alpha_1)$. Hence, the energy gain rapidly drops for $p \gg p_{\rm m1}$.
The behaviour of $e^{-\Gamma_1}$ is plotted in the left panel of Figure~\ref{fig:cutoff} where we show the impact of changing the diffusion coefficient from Kolmogorov to Bohm: for smaller momenta the asymptotic expression is the same as for a standard plane-parallel shock, but the way that such asymptotic value is approached depends on the spectrum of turbulence, being much more gradual for a Kolmogorov spectrum than for the case of Bohm diffusion.

While $\Gamma_1$ depends on the upstream, $\Gamma_2$ depends only on downstream quantities and produces a cutoff due to particle escape from the bubble boundary. The typical momentum, $p_{\rm m2}$, above which particles can escape the bubble efficiently is defined by the condition $\alpha_2=1$.
In the general case of spatially uniform  diffusion $D_2(p) = \kappa_2 p^{\delta_2}$, such a condition gives
\beq
  p_{\rm m2} = \left[ \frac{u_2 R_s}{\kappa_2} 
  \left(1 - \frac{R_s}{R_b}\right) \right]^{1/\delta_2} \,.
\eeq
The behaviour of $e^{-\Gamma_2}$ is shown in the right panel of Figure~\ref{fig:cutoff}, again for Bohm and Kolmogorov diffusion. Also in this case the Kolmogorov diffusion results in a broader cutoff with respect to the Bohm one but the behaviour for $p \ll p_{\rm m2}$ is basically identical. 

\begin{figure}
\centering
\includegraphics[width=.49\textwidth]{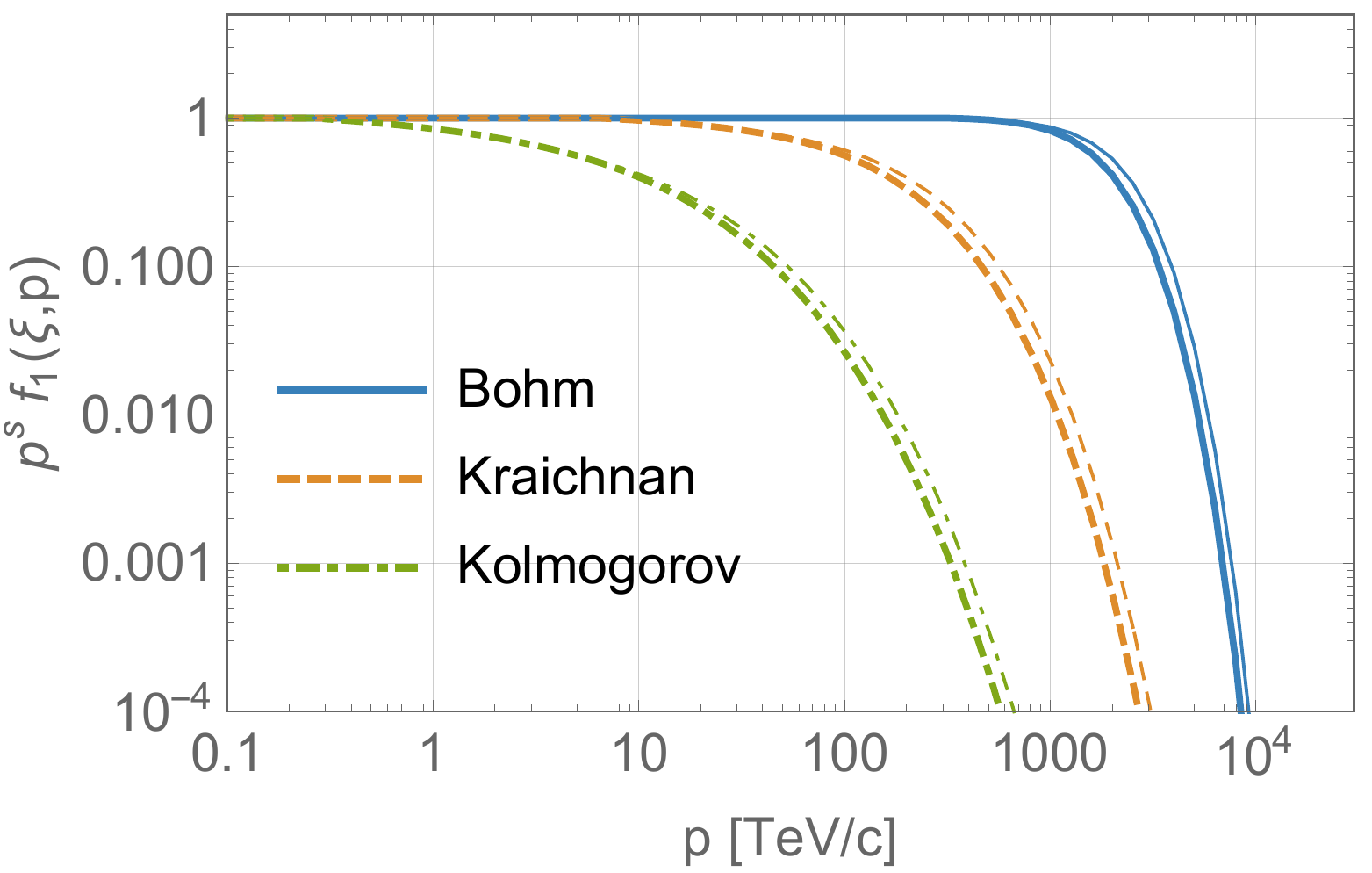}
\caption{Thick lines: distribution function of CR at the shock for different diffusion coefficients. Thin lines: corresponding escaping flux. The results refer to the benchmark case described in the text.}
\label{fig:fs_several_D}
\end{figure}

The effective maximum energy of accelerated particles is close to the smaller value between $p_{\rm m1}$ and $p_{\rm m2}$ and depends quite strongly on the type of diffusion at those energies. As an example, in Figure~\ref{fig:fs_several_D} we plot the full solution for $f_s$ assuming Kolmogorov, Kraichnan and Bohm diffusion. All cases are evaluated using typical parameters' values of a massive star cluster, namely: $\dot{M} = 10^{-4} M_{\odot}$, $v_w= 3000$~km s$^{-1}$, $t_{\rm age} = 10$~Myr,  $n_0 = 1$~cm$^{-3}$ and $\xi_{\rm CR} = \eta_{B} = 0.1$. In addition for the Kraichnan and Kolmogorov cases we fixed the turbulence injection scale at $L_c= 2$~pc.
One can see that an effective maximum energy up to $\sim$PeV can be reached only for Bohm diffusion while for the Kolmogorov case $p_{\max} \sim 10$\,TeV. The Kraichnan case is somewhat intermediate.

\begin{figure}
\centering
\includegraphics[width=.49\textwidth]{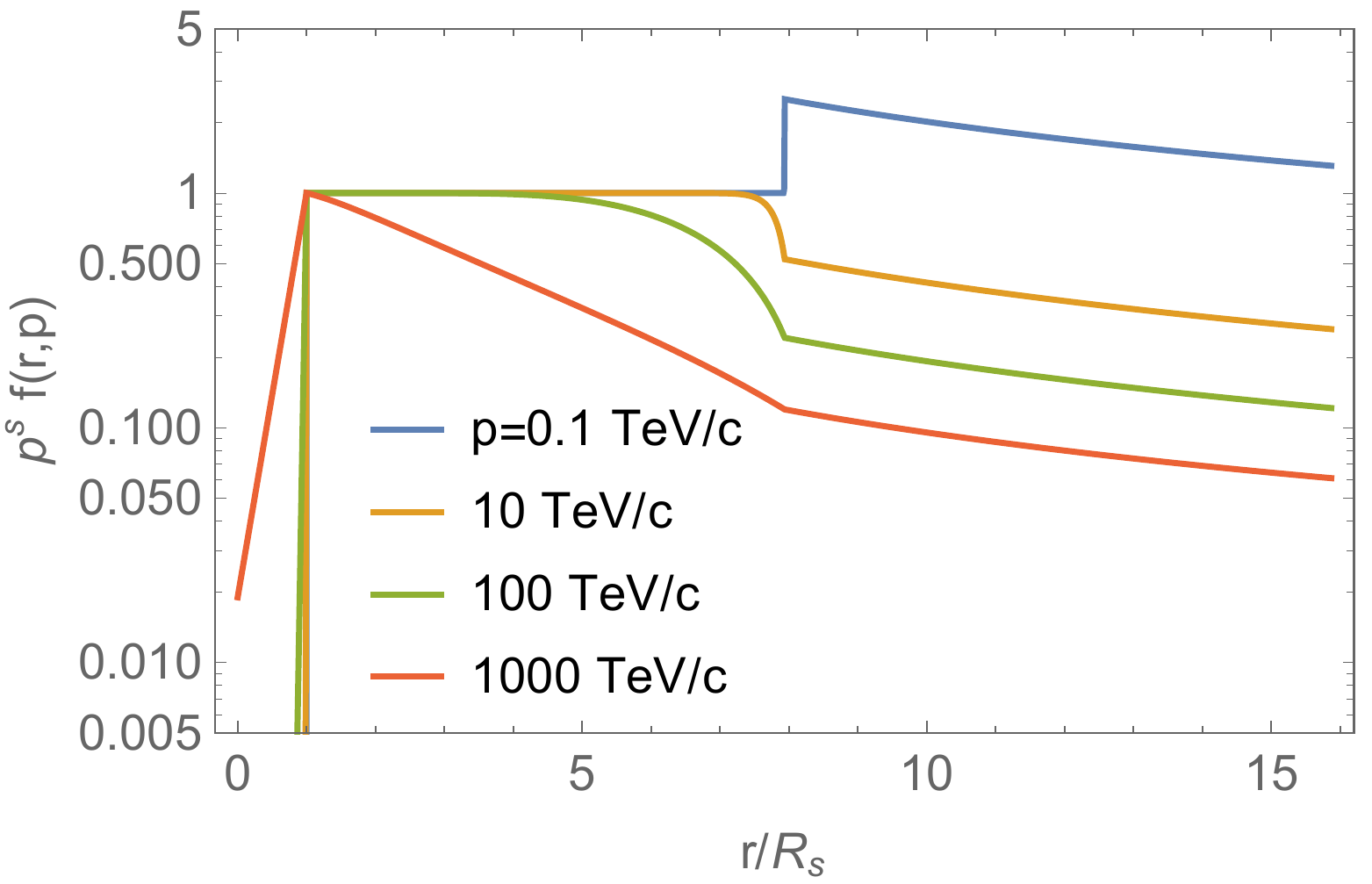}
\includegraphics[width=.49\textwidth]{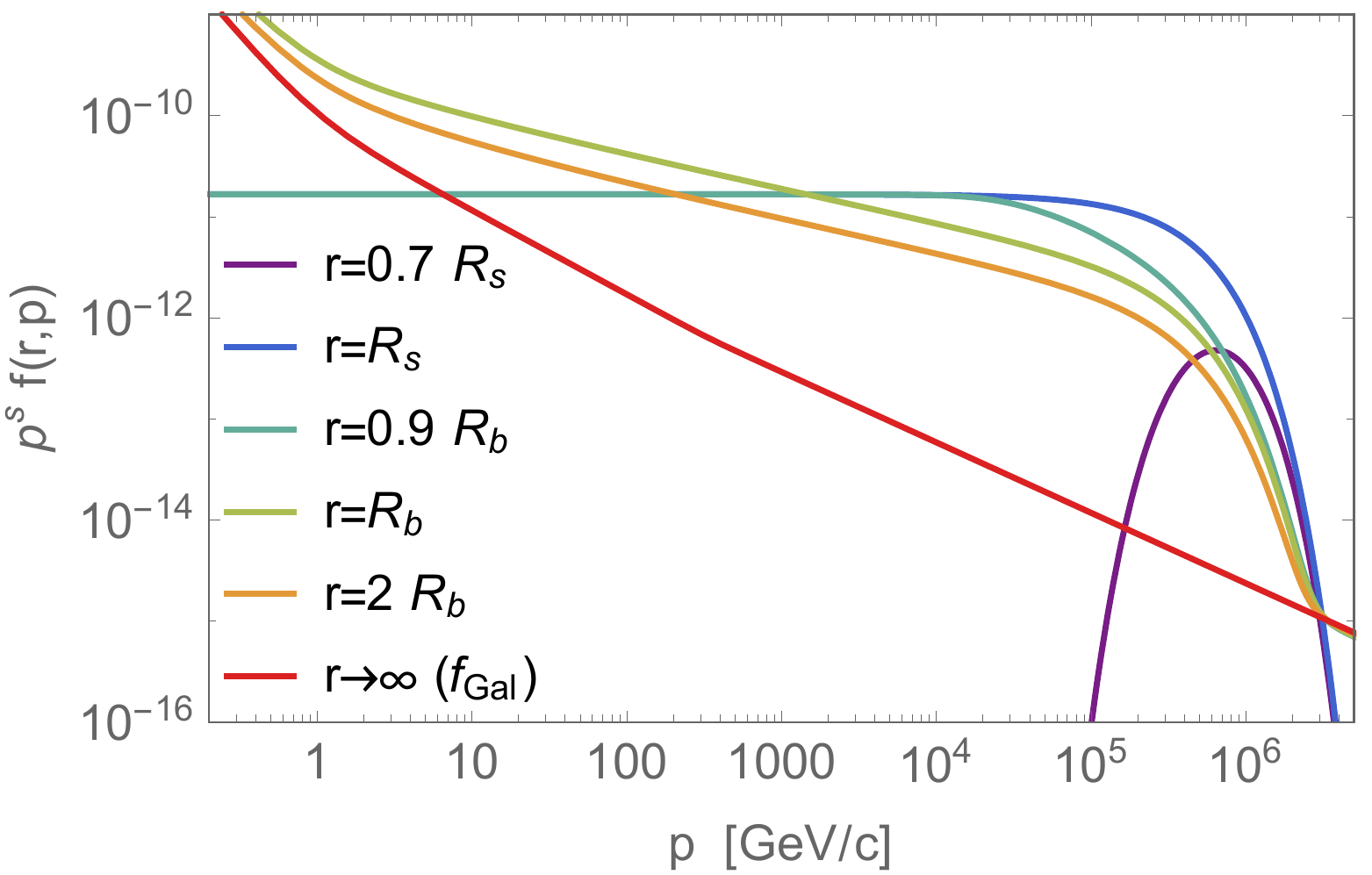}
\caption{Left panel: spatial profile of distribution function for different momenta normalized in such a way that $p^s f_s(R_s,p)=1$ for all $p$. Right panel: $p^s f_s(r,p)$ as a function of momentum for different position inside and outside the bubble.}
\label{fig:f_profile}
\end{figure}

We finally discuss the spatial behaviour of the solution \eqref{eq:f1}-\eqref{eq:f3}. Figure~\ref{fig:f_profile} shows the particle distribution function multiplied by $p^s$ as a function of position and momentum. As expected, $f$ is almost flat inside the shocked bubble for $p\ll p_{\rm m2}$ because at low energies the transport is dominated by advection. This is a general behaviour in sense that it would be valid even for the case of particles accelerated in the very center of the cluster provided that the diffusion time, $t_{\rm diff} \sim R_b^2/D_2$ is much smaller than the advection time $t_{\rm adv} \sim R_b/u_2$. Such a condition translate into $D_2 \ll u_2 R_b \approx 1.5 \times 10^{28} u_1/(2000\, {\rm km/s}) R_b/({100 \,\rm pc})$.
Inside the cold wind region, the particle distribution is strongly suppressed except at energies close to $p_{\rm m1}$. In the outer region, instead, $f\propto r^{-1}$ (see Eq.\eqref{eq:f3}) and slowly converge to $f_{\rm gal}$. In fact at $r=2 R_b$ the distribution function around the maximum energy is still two order of magnitude larger than $f_{\rm gal}$.

An important point to stress is the relative normalization between the spectrum accelerated at the TS and the Galactic one. To produce Figure~\ref{fig:f_profile} we have assumed $f_{\rm gal}$ equal to the local CR spectrum as inferred from the AMS-02 data \cite{AMS-P} (extrapolated up to $\sim 1$\,PeV) while the normalization of $f_{s}$ is obtained assuming that $\sim10\%$ of the wind kinetic luminosity is converted into CR luminosity. One can see from the right panel of Figure~\ref{fig:f_profile} that $f_{\rm gal}> f_s$ for $p\lesssim 10$\,GeV even if the acceleration efficiency is quite large. This happens because the flux of accelerated particles is a fraction of the wind flux, which at the TS has a density much smaller than the external ISM medium. In fact, using $\rho(R_s) = \frac{\dot M}{4\pi R_s^{2} v_{w}}$ we can write  $\rho(R_s)/\rho_0 \simeq 1.4 \times 10^{-4} \dot{M}_{-4}^{-2/5} v_{8}^{-6/5} \rho_{1}^{-2/5} t_{10}^{-2/5}$. 
Moreover, the CR density immediately outside the bubble is larger than its value inside the bubble at energies smaller than few hundreds of GeV.
This happen because particles are advected more efficiently from the bubble with respect to how efficiently they escape the circum-bubble region, hence they accumulate on the bubble boundary up to the moment when the two fluxes becomes equal. This fact could translate into a peculiar morphology of the corresponding hadronic $\gamma$-ray emission from the bubble that will be investigated in future works.

\section{Conclusions}
We have presented the first self-consistent model for particle acceleration at the termination shock of stellar winds. We have shown that the peculiar reverse geometry (with respect to, e.g., SNRs) and the spherical symmetry play an important role in determining the maximum energy. The latter is limited by two different conditions: the drop of energy gain for particles able to diffuse back to the cluster centre and the escape from the outer bubble boundary. 
In addition we have shown that the cutoff depends in a non trivial way by the diffusion properties inside the bubble. Applying this model to powerful stellar clusters, we showed that maximum energies up to $\sim$\,PeV can be reached assuming the the diffusion is close to be Bohm-like. Hence, it is of paramount importance to understand the kind of magnetic turbulence that develops in a wind-bubble structure in order to determine its diffusion properties and, in turn, its maximum energy.

\bibliographystyle{aa}
\bibliography{biblio}

%

\end{document}